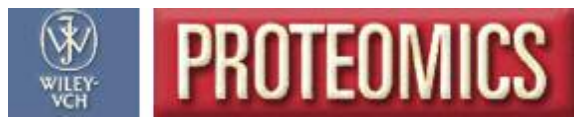

**Is the intrinsic disorder of proteins the cause of the scale-free architecture of protein-protein interaction networks?**

SHORT COMMUNICATION

# Is the intrinsic disorder of proteins the cause of the scale-free architecture of protein-protein interaction networks?

Santiago Schnell, Santo Fortunato and Sourav Roy

Indiana University School of Informatics, 1900 East Tenth Street, Eigenmann Hall, Bloomington, IN 47406, USA

---

**Abstract**

In protein-protein interaction networks certain topological properties appear to be recurrent: networks maps are considered scale-free. It is possible that this topology is reflected in the protein structure. In this paper we investigate the role of protein disorder in the network topology. We find that the disorder of a protein (or of its neighbors) is independent of its number of protein-protein interactions. This result suggests that protein disorder does not play a role in the scale-free architecture of protein networks.

---

**Correspondence:** Santiago Schnell, Indiana University School of Informatics and Biocomplexity Institute, 1900 East Tenth Street, Eigenmann Hall 906, Bloomington, IN 47406, USA. Email: schnell@indiana.edu. Fax: +1-812-856-1955.

There is much current interest in the structure of biological networks, in particular focusing on their topology [1]. Certain organizational properties appear to be a recurrent feature of biological systems. A power-law probability distribution for the number of links adjacent to the same node (degree) has been reported in systems as diverse as protein folding [2], networks of metabolic reactions [3-4] and ecological food webs [5]. A power-law degree distribution indicates that connectivity is quite heterogeneous on the network, so that many nodes having just a few connections coexist with a considerable number of nodes with many connections (known as hubs). Scale-free networks are resilient to random failures but vulnerable to targeted attacks against the most connected elements or hubs [6]. Network topology is implicated in conferring robustness of network properties, for example in setting up patterns of cellular differentiation in gene regulatory networks in the early development of the fruit fly [7-8], or in providing resistance to mutation of environmental stress [9-10]. Therefore, the determination of the topology of a biological network has become important for assessing the stability, function, dynamics and design aspects of the network [11].

Network topologies are determined using a variety of sample methods, which are then used to infer the topology of the whole network [12]. The application of these methodologies to protein-protein interaction (PPI) networks has revealed that they are scale-free [13-18]. This architecture of the PPI networks must be reflected in the protein structure [19-20]: What is the structural feature of hub proteins which allows them to interact with a large number of partners?

Recently, Fernández and Berry [21] made an attempt to answer this question. They examined the structure wrapping, which protects a protein from water attack, in the PPI network of *Saccharomyces cerevisiae* [13]. They found that relaxing the structure of the packing interface increases the number of binding interactions in a protein. However, this study did not examine the correlation between the degree of a protein (number of interactions) in a PPI network and the extent of its overall wrapping deficiency. As a consequence, the results of this study are of limited applicability for understanding the architecture of a PPI network.

Fernández and Berry [21] have found that there is a direct correlation between the number of hydrogen bonds that are deficiently protected from water and the inherent structural disorder of a protein. Intrinsically disordered proteins do not have specific three-dimensional structures [22-23]. They exist in wide-motion conformation ensembles both *in vitro* and *in vivo* [24-25]. The intrinsically disordered regions of a protein can provide flexible linkers between functional domains, which facilitate binding diversity and can explain the existence of protein hubs [19-20]. Dunker and co-workers [19] have proposed two roles for intrinsic disorder in PPI:

      **I**. Intrinsic disorder can serve as the structural basis for hubs promiscuity, and

      **II**. Intrinsically disordered proteins can bind to structured hub proteins.

In this short communication, we determine if there is a correlation between the number of interactions of a protein in a PPI network and the disorder of the protein itself or that of its neighbors.

We started downloading the PPI networks for *Homo sapiens* (human), *Saccharomyces cerevisiae* (yeast), *Drosophila melanogaster* (fly), *Caenorhabditis elegans* (worm) and *Escherichia coli* (bacterium) from the Biomolecular Interaction Network Database (BIND http://bind.ca) [26]. The information collected was cured using the low-throughput track of the database and by filtering out redundant interactions. First, we carried out a standard analysis of the five networks under consideration. In Table 1 we list the main attributes of the networks: the number of nodes, the number of links, the average degree and the diameter. The average degree reveals that all networks are quite sparse; as a matter of fact, they are fragmented in many components, with most components consisting of a single node. Due to the limited size of the networks, a statistical analysis is meaningful only for the largest systems. Figure 1 displays the degree distribution of the human PPI network. The data points were obtained by averaging the probability within logarithmic bins of degree, to reduce fluctuations. The distribution looks quite skewed, with a tail following approximately a power law with exponent 2.5.

Our aim is to address quantitatively the issue of the existence of a correlation between the topological connectivity of a protein, expressed by its degree k, and the disorder of the protein itself or that of its neighbors. We employ the VL3 model [27] to predict intrinsically disordered regions of protein sequences for the PPI networks under consideration. The VL3 predictor is a feed-forward neural network algorithm which uses attributes such as amino acid compositions, average flexibility [28] and sequence complexity [29] for calculating the disorder score. The disorder score is a value between 0 and 1. We calculate the number of disordered regions in a protein as those putative segments which were of 10 or more residues in length with a disorder score of 0.5 or greater. We estimate these segments, because the increasingly recognized mechanism of interaction between intrinsically disordered proteins and their binding partners is through molecular recognition elements. These elements are typically located within disordered regions, which provide enough physical space and flexibility to accommodate access of a variety of potential partners thereby facilitating binding with multiple and diverse targets [30-32].

We computed the standard Pearson correlation coefficient between the degree of all proteins of a network and their disorder, as expressed by the disorder score and by the number of disordered regions. The corresponding correlation coefficients for the protein (node) degree versus disorder score and protein (node) degree versus number of disordered regions are listed in Table 2. They are all close to zero. This result indicates the absence of correlations between the number of interactions of a protein and its disorder coefficient, and between the number of interactions of a protein and its number of disordered segments. We completed the analysis by calculating the Pearson correlation coefficients between the degree k of a protein and the average disorder of its neighbors. The corresponding coefficients are listed in Table 3, and they are again very close to zero, hinting at the absence of correlations between the degree of the protein and the average disorder of its neighbors.

We have as well investigated the relation between degree and disorder, as the knowledge of the correlation coefficient only gives a global indication on the correlation between two variables. This type of analysis is particularly suitable for large systems, as one ideally wishes to have a significant number of data points for each degree value. Therefore we present here the result of the analysis for the largest network we have, i.e. the human PPI network. Figure 2 shows four scatter plots between degree and disorder. To show the trend of the plots, we averaged disorder within logarithmic bins of degree. The resulting curves are essentially flat, which shows that the disorder of a protein (or of its neighbors) is independent of its degree, confirming the absence of correlation we had derived from the examination of the correlation coefficients.

From our analysis, we find that an increase in the disorder score or number of disordered regions of a protein does not increase its topological connectivity (nor does the disorder of its neighbors). We also learned that the average disorder of the neighbors of a hub protein does not correlate with the number of interactions of the hub protein. A limitation of our analysis is that we have calculated the disorder coefficient for the entire protein sequence, and we do not specifically study the regions involved directly in protein-protein binding. This investigation could be refined by studying the disorder of the residuals involved in the binding of proteins. The intracellular environment is heterogeneous, structured and compartmentalized. Biochemical reactions are diffusion limited and proteins are contained in different compartments [33]. The networks considered in this communication were treated as a homogeneous landscape, because they do not reflect the underlying cellular and molecular process. Whether this would also affect the results of our analysis needs to be determined. By itself, the discovery that PPI networks have scale-free properties is of limited use to biologists [34]. An important research direction is investigating what structural or molecular properties of the proteins are the bases for the scale-free architecture of PPI networks.

*We are grateful for the useful suggestions made by Profs. A. Keith Dunker and Vladimir N. Uversky (Department of Biochemistry and Molecular Biology, Indiana University School of Medicine), and Profs. Predrag Radivojac (Indiana University School of Informatics) during the development of this work.  We would like to thank Alessandro Vespignani (Indiana University School of Informatics) for his critical comments. The VL3 predictor for protein disorder was kindly provided Prof. Radivojac. SS would like to the FLAD Computational Biology Collaboratorium at the Gulbenkian Institute in Oeiras, Portugal, for hosting and providing facilities used to conduct part of this research between April 2nd-10th. This research has been supported by the Division of Information & Intelligent Systems, National Science Foundation, Grant No. 0513650. Any opinions, findings, conclusions or recommendations expressed in this paper are those of the authors and do not necessarily reflect the views of the National Science Foundation or the United States Government.*

**References**

[1] Crampin, E.J., Schnell, S., *Prog. Biophys. Mol. Biol.* 2004, *86*, 1-4.

[2] Koonin, E.V., Wolf, Y.I., Karev, G.P., *Nature* 2002, *420*, 218–223.

[3] Jeong, H., Tombor, B., Albert, R., Oltvai, Z.N., *et al.*, *Nature* 2000, *407*, 651–654.

[4] Wagner, A., Fell, D.A., *Proc. R. Soc. London Ser. B* 2001, *268*, 1803–1810.

[5] Solé, R.V., Ferrer-Cancho, R., Montoya, J.M., Valverde, S., *Complexity* 2003, *8*, 20–33.

[6] Albert, R., Jeong, H., Barabási, A.-L., *Nature* 2000, *406*, 378-382.

[7] von Dassow, G., Meir, E., Munro, E.M., Odell, G.M., *Nature* 2000, *406*, 188–192.

[8] Albert, R., Othmer, H.G., *J. Theor. Biol.* 2003, *223*, 1–18.

[9] Wagner, A., *Nat. Genet.* 2000, *24*, 335-361.

[10] Yeong, H., Mason, S.P. Barabási, A.-L., Oltvai, Z.N., *Nature* 2001, *411*, 41-42.

[11] Barabási, A.-L., Oltvai, Z.N., *Nature Rev. Genet.* 2004, *5*, 101–113.

[12] Albert, R., Barabási, A.-L., *Rev. Mod. Phys.* 2002, *74*, 47-97.

[13] Uetz, P., Giot, L., Cagney, G., Mansfield, T.A., *et al.*, *Nature* 2000, *403*, 623-627.

[14] Ito, T., Chiba, T., Ozawa, R., Yoshida, M., *et al.*, *Proc. Natl. Acad. Sci. USA* 2001, *98*, 4569-4574.

[15] Reboul, J., Vaglio, P., Rual, J.F., Lamesch, P., *et al.*, *Nat. Genet.* 2003, *34*, 35-41.

[16] Giot, L., Bader, J.S., Brouwer, C., Chaudhuri, A., *et al*., *Science* 2003, *302*, 1727-1736.

[17] Li, S.M., Armstrong, C.M., Bertin, N., Ge, H., *et al*., *Science* 2004, *303*, 540-543.

[18] Han, J.D.J., Bertin, N., Hao, T., Goldberg, D.S., *et al.*, *Nature* 2004, *430*, 88-93.

[19] Dunker, A.K., Cortese, M.S., Romero, P., Iakoucheva, L.M., *et al.*, *FEBS Journal* 2005, *272*, 5129-5148.

[20] Uversky, V.N., Oldfield, C.J., Dunker, A.K., *J. Mol. Recog.* 2005, *18*, 343-384.

[21] Fernández, A., Berry, R.S., *Proc. Natl. Acad. Sci. USA* 2005, *101*, 13460-13465.

[22] Wright, P.E., Dyson, H.J., *J. Mol. Biol.* 1999, *293*, 321-331.

[23] Dunker, A.K., Lawson, J.D., Brown, C.J., Williams, R.M., *et al.*, *J. Mol. Graph. Model.* 2001, *19*, 26-59.

[24] Uversky, V.N., Gillespie, J.R., Fink, A.L., *Proteins* 2000, *41*, 415-427.

[25] Dedmon, M.M., Patel, C.N., Young, G.B., Pielak, G.J., *Proc. Natl. Acad. Sci. USA* 2002, 99, 12681–12684.

[26] Bader, G.D., Donaldson, I., Wolting, C., Ouellette, B.F.F., *et al.*, *Nucleic Acids Res.* 2001, *29*, 242-245.

[27] Obradovic, Z., Peng, K., Vucetic, S., Radivojac, P., *et al.*, *Proteins*, 2003, *53*, 566-572.

[28] Vihinen, M., Torkkila, E., Riikonen, P., *Proteins* 1994, *19*, 141-149.

[29] Wootton, J.C., *Comput. Chem.* 1994, *18*, 269-285.

[30] Fuxreiter, M., Simon, I., Friedrich, P., Tompa P., *J. Mol. Biol.* 2004, *338*, 1015-1026.

[31] Oldfield, C.J., Cheng, Y., Cortese, M.S., Romero, P., *et al.*, *Biochemistry* 2005, *44*, 12454-12470.

[32] Roy, S., Schnell, S., Radivojac, P., Comp. Biol. Chem. 2006, DOI: 10.1016/j.compbiolchem.2006.04.005

[33] Schnell, S., Turner, T.E., *Prog. Biophys. Mol. Biol.* 2004, *85*, 235-260.

[34] Bray, D., *Science* 2003, *301*, 1864-1865.

Table 1. Protein-protein interaction network properties.

| Species | Nodes | Links | Average degree | Diameter |
|---|---|---|---|---|
| Human | 2,758 | 3,237 | 2.347 | 27 |
| Yeast | 881 | 1,107 | 2.513 | 21 |
| Fly | 353 | 280 | 1.586 | 8 |
| Worm | 244 | 244 | 2.000 | 7 |
| Bacterium | 166 | 132 | 1.590 | 4 |

Table 2. Pearson correlation coefficients between the degree of a protein and its disorder: the latter is expressed by the disorder score or by the number of disordered regions. The degree of a node corresponds to number of interactions of a protein. The disorder coefficients are calculated using the VL3 predictor [27].

| Species | Correlation Coefficient k-disorder | Correlation Coefficient k-disordered regions |
| --- | --- | --- |
| Human | -8.45212E-004 | 1.75295E-002 |
| Yeast | 3.80361E-002 | 7.76235E-002 |
| Fly | 7.10621E-003 | 2.88277E-002 |
| Worm | -1.18723E-002 | -6.00492E-002 |
| Bacterium | 5.15177E-002 | 0.11902 |

<u>Table 3</u>. Pearson correlation coefficients between the degree of a protein and the average disorder of its neighbors: the latter is expressed by the disorder score or by the number of disordered regions. The degree of a node corresponds to the number of interactions of a protein. The disorder coefficients are calculated using the VL3 predictor [27].

| Species | Correlation Coefficient k-disorder | Correlation Coefficient k-disordered regions |
|---------|-----------------------------------|---------------------------------------------|
| Human | 2.27327E-002 | 4.00298E-002 |
| Yeast | -1.63826E-002 | 2.15817E-002 |
| Fly | 1.59666E-002 | -1.09563E-002 |
| Worm | 0.26913 | -1.15449E-002 |
| Bacterium | 7.39843E-002 | 0.11434 |

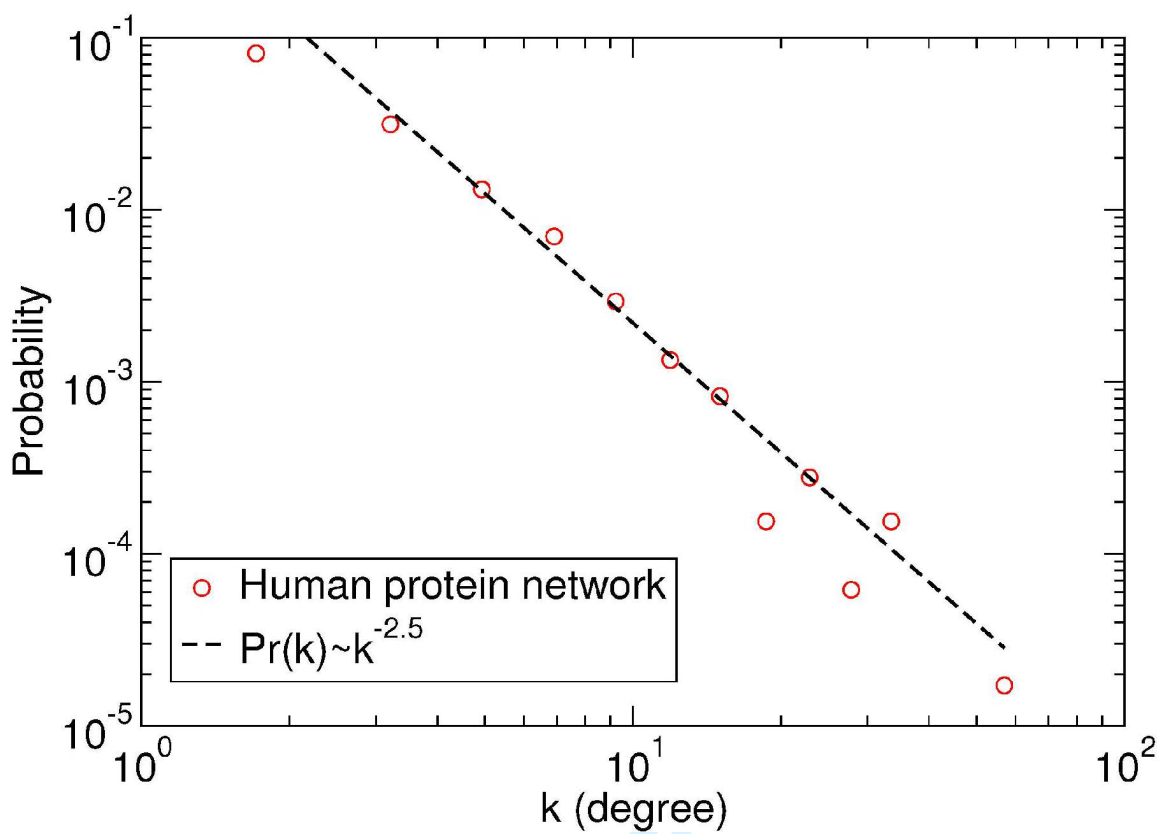

<u>Figure 1</u>. Degree distribution for the human protein-protein interaction network. The degree of a node corresponds to the number of interactions of a protein. Note that the distribution is quite broad, spanning two orders of magnitude in degree. The dashed line is a tentative power law fit of the tail of the distribution: the exponent is 2.5.

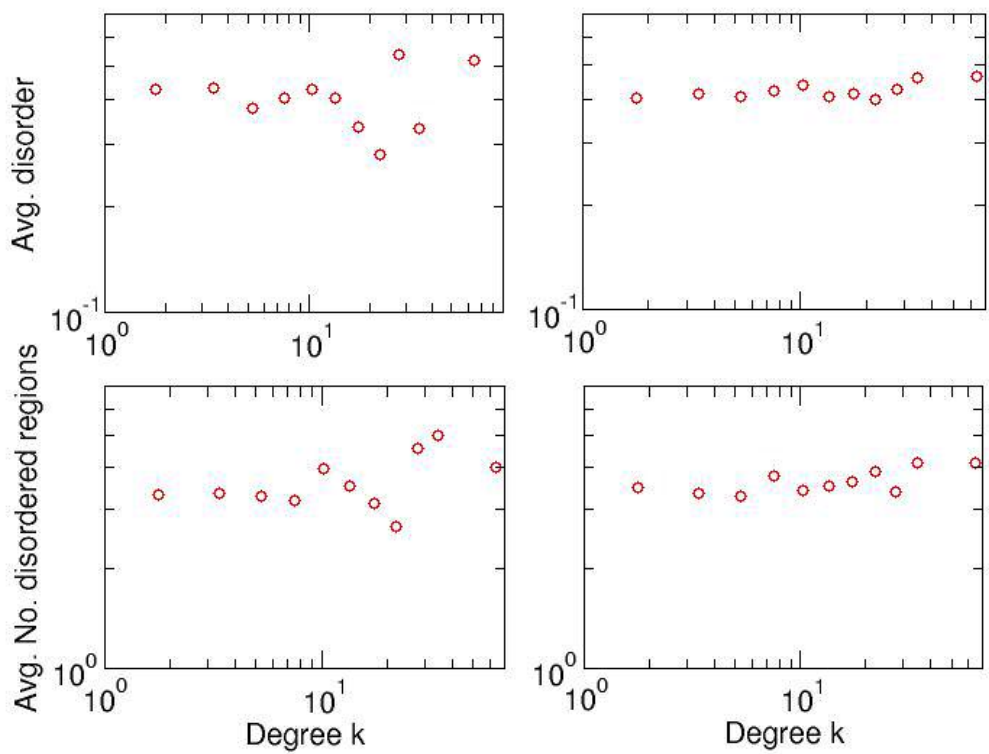

<u>Figure 2</u>. Relation between disorder and degree for the human PPI network. The two plots on the left show the variability with degree of the disorder of a protein, with the latter being expressed by the disorder score (top left) and by the number of disordered regions (bottom left). The two plots on the right show how the average disorder of the neighbors of a protein varies with the degree of the protein. Again, we use both the disorder score (top right) and the number of disordered regions (bottom right).